\documentclass{IEEEtran}

\usepackage{multirow}
\usepackage{graphicx}
\usepackage{amsmath}
\usepackage{dblfloatfix}  
\usepackage{caption}
\usepackage{cite}
\usepackage{graphicx}
\usepackage{amsmath}
\usepackage{algorithm}
\usepackage{algorithmic}
\usepackage{booktabs}
\usepackage{comment}
\usepackage[symbol]{footmisc}
\renewcommand{\thefootnote}{\fnsymbol{footnote}}
\usepackage[symbol]{footmisc}
\usepackage[utf8]{inputenc}
\usepackage{longtable}

\usepackage{xcolor}
\definecolor{R1}{rgb}{1 0 0}
\definecolor{B1}{rgb}{0 0 1}

\newcommand{\REV}[1]{#1}
\newcommand{\REVNEW}[1]{#1}
\newcommand{\SREV}[1]{}

\usepackage{soul}
\usepackage{cancel}
\usepackage[normalem]{ulem}
\def\BibTeX{{\rm B\kern-.05em{\sc i\kern-.025em b}\kern-.08em
    T\kern-.1667em\lower.7ex\hbox{E}\kern-.125emX}}

\begin{document}

\title{The Systematic Design of Non-commensurate Impedance Matching Tapers for Ultra Wideband Gradient-Index (GRIN) Lens Antennas}


\author{Wei Wang,~\IEEEmembership{Student Member,~IEEE}, Nicolas Garcia,~\IEEEmembership{Student Member,~IEEE}, Jonathan Chisum, \IEEEmembership{Senior Member,~IEEE}\vspace{-0.9cm}%
\thanks{Manuscript received August 28, 2020; revised February 13, 2021; revised April 28, 2021; accepted June 5, 2021. Date of publication ??; date of current version June 9, 2021. This work supported in part by the Department of the Navy, Office of Naval Research under contract N00014-20-C-1067. \textit{(Corresponding author: Jonathan Chisum)}}
\thanks{The authors are with the Department of Electrical Engineering, University of Notre Dame, Notre Dame, IN, 46556 USA. E-mail: jchisum@nd.edu, wwang23@nd.edu}
\thanks{Color versions of one or more of the figures in this paper are available online at http://ieeexplore.ieee.org.}
\thanks{Digital Object Identifier ??.????/???.????.???????}
}
\markboth{IEEE TRANSACTIONS ON ANTENNAS AND PROPAGATION, VOL. XX, NO. YY, ZZZ 2021}%
{}

\IEEEtitleabstractindextext{%
\begin{abstract}
\REV{We propose a general method for designing wideband matching tapers in inhomogeneous media where phase velocity is coupled to the taper impedance profile. Such tapers are used to match wideband gradient index (GRIN) lens antennas. To simplify fabrication tapers are often constrained to physically uniform layers wherein commensurate line theory cannot predict the frequency response. Therefore, we present a new design algorithm which derives an effective permittivity $\epsilon_{eff}$ to equalize the electrical length of commensurate and non-commensurate line tapers. The algorithm provides a systematic design method with predictable frequency response for non-commensurate line tapers. Nevertheless, there are several unavoidable nonidealities present in such discretized tapers which we discuss and provide recommendations for mitigation. The algorithm is used to design a Klopfenstein taper with return loss better than 15\,dB from 8 to 78\,GHz. The design is fabricated and measurements agree with simulation across the WR90, WR28, and WR12 bands. An approximate efficiency formula is proposed which predicts aperture efficiency of taper-matched lenses without the need for time-consuming full-wave simulations. Various lenses are designed and compared to highlight the advantages of Klopfenstein tapers in GRIN lens design. The results demonstrate the usefulness of the proposed design method.} 
\end{abstract}

}

\maketitle

\IEEEdisplaynontitleabstractindextext

\IEEEpeerreviewmaketitle

\section{Introduction}

\IEEEPARstart{W}{e} present a method for the design of broadband impedance matching tapers constrained to uniform layer thickness for application in quasi-optical systems such as gradient index (GRIN) lens antennas. GRIN lenses provide a low-profile means of realizing high aperture efficiency over wideband operation but only if they are properly impedance matched over the band \cite{Garcia_Matched_2020}. \REV{Because of the wide variety of applications which use GRIN lenses, their matching sections should be readily designed across many frequency bands (from Ku-band to W-band) \cite{Garcia_Matched_2020,MaCui2_GRIN_2019,Mahmoud_WbandGRIN_2014,Imbert_LTCClens_2017} and with a variety of performance metrics. For example, cost and size-constrained Ku/Ka-band commercial satellite communications demand large bandwidth and high aperture efficiency to enable compact and high-gain apertures \cite{Yahya_sateillite_2015}. Such applications might prefer low cutoff frequency over passband match level. In contrast, high power, wideband radar and communications systems\cite{Sun_highpower_2017} prefer extremely low passband reflections to reduce power reflected back toward the source. In polarization sensitive applications very low passband reflections might be required in order to maintain a nearly equal passband transmission efficiency for TE- and TM-modes\cite{Su_Polarization_2019}. These demands require a flexible and predictable matching solution for GRIN lenses.} 

\begin{figure}[tb]
    \centering
    \includegraphics[width=9cm]{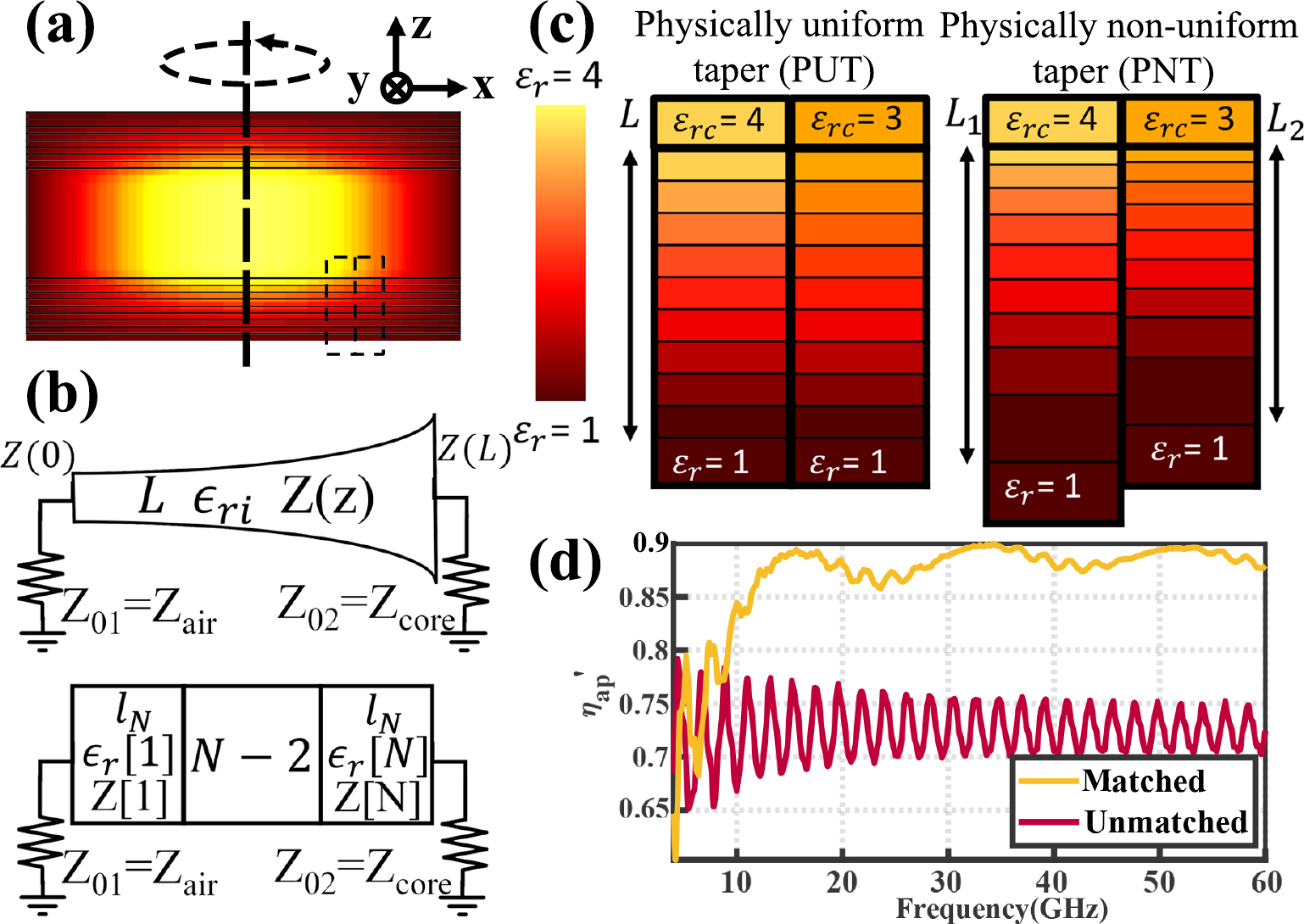}
    \caption{(a) Symmetrical GRIN lens cross section and permittivity profile with matching tapers (vert./horiz. axes not to scale). (b) Ideal, continuous taper $Z(z)$ and discretized PUT $Z[n]$. (c) PUT and PNT in adjacent unit-cells with core permittivity $\epsilon_{rc}$. \REV{(d) Normalized aperture efficiency $\eta_{ap}'$ (see (\ref{eq:ApPrime})) for unmatched and Klopfenstein taper GRIN lenses.}}
    \label{fig:structure}
    \vspace{-0.25cm}
\end{figure}

For practical considerations, the 3D permittivity profile required to implement a GRIN lens is often realized on a uniform grid as in \cite{Anastasios_GRIN2020,Su_PCBGRIN_2018,MaCui2_GRIN_2019,MaCui1_rings_2009,Sun_highpower_2017,MaCui3_APL_2011,Mahmoud_WbandGRIN_2014,Imbert_LTCClens_2017,Elef_Matched_2018,Garcia_Matched_2020}. \REV{As we will show below, this constraint makes it impossible to reliably predict the performance of an impedance matching taper using classical commensurate-line taper theory \cite{Pozar05Mwave}.} The general symmetrical GRIN lens cross section and permittivity distribution considered in this work is shown in Fig.\,\ref{fig:structure}(a). In such a configuration the GRIN lens comprises a stack of layers, each with a unique in-plane ($xy$-plane) permittivity profile. While, in general, the thickness of each layer may be independent from other layers, in order to practically realize a continuous 3D index gradient the GRIN profile is often discretized on a uniform grid. As such, the tapers in this work are constrained to be of uniform layer thickness $l_N$ along the $z$-axis.

Figure\,\ref{fig:structure}(b,top) shows a simplified taper in which a continuously varying characteristic impedance, $Z(z)$, is realized by varying the width of the transmission line. In order for such a taper to serve as the surface impedance matching section of a GRIN lens (dashed rectangles of Fig.\,\ref{fig:structure}(a)), \REV{$Z(z), z\in[0,L]$} is usually fabricated as $N$ discrete segments as shown in Fig.\,\ref{fig:structure}(b,bottom). For transmission lines in which impedance can be set by cross section geometry (e.g., stripline) the phase velocity may remain constant for each segment and thus commensurate line (electrically uniform) tapers are also physically uniform tapers (PUT). \REV{In contrast, tapers realized with transmission lines in which the phase velocity and characteristic impedance are simultaneously modified through the refractive index of the medium (e.g., microstrip or quasi-optical systems) should be physically non-uniform tapers (PNT) to maintain the desired performance characteristics of the taper.} For planar transmission lines such as microstrip, it is simple to realize the required PNT by adjusting layer thickness \cite{KuesterHsu_Tapers}, but for quasi-optical matching tapers which prefer a PUT realization, modifying the permittivity to realize the impedance taper results in non-commensurate lines (electrically non-uniform) and the overall response deviates from the intended design \cite{Blomfield1983noncommensurate}. Therefore this work presents a method for maintaining the key performance parameters of an impedance matching taper realized with non-commensurate lines.

Figure\,\ref{fig:structure}(c) shows detail of two hypothetical adjacent GRIN lens unit-cells from Fig.\,\ref{fig:structure}(a), which provide a match from free-space ($\epsilon_r=1$) to a radial varying core permittivity, here $\epsilon_{rc}=4$ and $\epsilon_{rc}=3$. Each unit-cell is shown in a PUT realization and a PNT realization. The comparison shows PNT layer thickness varies with the layer permittivity and its overall thickness depends on the core permittivity, but the PUT layer thickness\footnote{layer ``thickness'' and layer ``length'' will be used interchangeably throughout.} is uniform and the overall thickness is constant making PUTs convenient for fabrication of GRIN lenses. \REV{Figure\,\ref{fig:structure}(d) shows the approximate aperture efficiency (see (\ref{ap_etaT})) of such a lens with and without the proposed matching sections---the PUT-matched lens, described in Section\,\ref{sec:Demo:Comparison} below, is simple to fabricate and provides a significant increase in passband efficiency.}

Because of the benefit of PUTs in quasi-optical structures, we seek a design method for realizing PUTs. In section \ref{sec:theory} we present the theory and propose an algorithm for their design. In section \ref{sec:impairment} we discuss the unavoidable nonidealities which result from such discretized and non-commensurate tapers and suggest methods for mitigating the impairments when possible. In section IV we present the design and measurement of a prototype taper section comprising nine physically uniform layers for use in GRIN lens antennas operating across the 8--78\,GHz band.\SREV{and then conclude with an example lens design using the proposed tapers and exhibiting a matched transmission efficiency of greater than 97\% in the passband above the taper cutoff frequency of 15\,GHz.} \REV{Section V presents a complete GRIN lens design procedure, proposes an approximate efficiency formula which is useful for guiding design, and provides a detailed comparison of Klopfenstein taper-matched GRIN lenses with other common matching approaches (e.g., exponential tapers, quarter wave sections).}

\section{Theory} \label{sec:theory}

\REV{Tapers provide an impedance match over a prescribed operating band which results in increased transmission efficiency and aperture efficiency. However, taper designs used in the recent literature \cite{Garcia_Matched_2020}\cite{biswas_high_2020}\cite{Lee11SIW} only consider the impedance profile $Z(z)$ and neglect the fact that impedance and phase constant are coupled through the permittivity of the medium. As stated above, if PNTs are used to match GRIN lenses the taper response can be predicted from well-established commensurate line theory \cite{Pozar05Mwave}. However, since nearly all realized GRIN lenses use discretized PUTs (owing to their reduced complexity of fabrication and design), they exhibit nonuniform electrical length of each taper layer which causes taper performance $\Gamma(\theta)$ to deviate from the corresponding ideal taper response. Indeed there is no predictive formula for the frequency response $\Gamma(\theta)$ of PUTs. The lack of a predictable response means tapers are designed approximately and therefore not optimally. This is especially problematic for Klopfenstein tapers which provide a design tradeoff between low-frequency cutoff and passband match---the tradeoff cannot be accurately studied without a predictable response.}


\REV{To develop a method which considers the coupling of impedance and phase velocity, we need to not only acquire the impedance profile but preserve the performance of the ideal taper in a PUT. Consider the ideal taper structure shown in} Fig.\,\ref{fig:structure}(b,top). TEM electromagnetic waves travel through the ideal taper with uniform phase velocity \SREV{$v_g$}\REV{$v_p$} while the local characteristic impedance $Z(z)$ varies according to the taper matching profile. Then the impedance profile $Z(z)$ of the taper (e.g., exponential, Klopfenstein) is discretized on a uniform grid of $N$ layers resulting in $Z[n]$. Now if each layer is realized with a uniform electrical length by varying the physical layer thickness we arrive at a PNT. \REV{If each layer is realized with a uniform physical length (and correspondingly, varying electrical length) we arrive at a PUT.} 
\REV{In order to derive a PUT with a predictable frequency response we present the following procedure, summarized in Algorithm\,\ref{alg:taper}. Note that while the algorithm is demonstrated with a Klopfenstein taper (because it provides a prescribed match level in the shortest possible length), other tapers could be used.} 

First consider the ideal Klopfenstein taper profile and response: for a given taper length $L$ and matching range $Z_{01}$ to $Z_{02}$ \REV{($Z_{01}$ $>$ $Z_{02}$)}, the Klopfenstein profile achieves a tradeoff between maximum passband ripple, $\Gamma_{max}$, and \SREV{low} cutoff frequency, $f_{c}$\footnote{\REV{As is standard for a Klopfenstein taper, the cutoff frequency is defined as the lowest frequency for which the input reflection coefficient is $\leq\Gamma_{max}$. It is not the frequency of the first null or the first sidelobe peak.}}, according to,

\begin{equation} 
\label{ga and band}
    \Gamma_{max} = \frac{\Gamma_0}{\cosh A},  A=\frac{2\pi f_{c}}{c_{0}}\REV{\epsilon_{eff}}L,
\end{equation}

\noindent where $\Gamma_0=\left(Z_{01}-Z_{02}\right)/\left(Z_{01}+Z_{02}\right) \approx 0.5\ln\ \left(Z_{01}/Z_{02}\right)$, $c_0$ is the speed of light in vacuum, and \REV{$\epsilon_{eff}$} is the homogeneous, uniform permittivity of an ideal continuous TEM Klopfenstein taper. With $Z_{01}$, $Z_{02}$, \REV{$\epsilon_{eff}$}, $\Gamma_{max}$ and $f_{c}$, one can derive the unique impedance profile and corresponding response of the ideal Klopfenstein taper \cite{Pozar05Mwave}.


Next, to realize a PNT 
\REV{the ideal taper profile is discretized with} $N$ layers of equal electrical length. For large $N$, assigning the impedance of each layer $Z[n]$ as the corresponding central value from the original continuous impedance profile $Z(z)$, results in a trivial modification of the ideal continuous taper response. \REV{For quasi-optical realizations with non-magnetic media} the permittivity value of the $n^{\textrm{th}}$ layer is $\epsilon_{r}[n]={(377/Z[n])}^2$ with phase velocity $v_{p}[n]= {c_0}/{\sqrt{\epsilon_{r}[n]}} $. To maintain commensurate layers, the physical thickness $l[n]$ can be found for each layer using $\epsilon_{r}[n]$, $n\in \left[1,N\right]$. These layers can be thought of as quarter wavelength transformers at a common central frequency $f_o$,
\begin{equation} \label{cf}
    f_{o} = \frac{c_0}{4\sqrt{\epsilon_{r}[n]}l[n]}.
\end{equation}
\noindent We note that the discretized, N-layer Klopfenstein PNT is a Chebyshev transformer for large $N$.

Finally, to realize a PUT
$N$ layers of equal physical thickness $L/N$ are used \REV{instead of $l[n]$}. Suppose $Z_{01},Z_{02},L,N,f_c$ are known: the general approach we pursue for realizing a PUT from the ideal taper impedance profile is to find the permittivity \REV{$\epsilon_{eff}$} of the ideal taper which, when discretized as a PUT results in equal total phase, $\theta_{PUT}=\theta_{\textrm{ideal}}$, and equal total thickness, $L_{PUT}=L_{\textrm{ideal}}$. These conditions taken together maintain the cutoff frequency $f_{c}$ which is considered the key parameter for quasi-optical components which are ideally thin relative to the lens diameter (and thus achieving a low $f_{c}$ is challenging). Equality of total phase requires,
\begin{equation} \label{thetaeqn}
    \sum_{n=1}^{N}{\frac{2\pi f_{c}}{c_0}\sqrt{\epsilon_{r}[n]}l_{N}}= \frac{2\pi f_{c}}{c_0}\sqrt{\REV{\epsilon_{eff}}}L.
\end{equation}

\noindent Notice that the uniform \REV{$\epsilon_{eff}$} of the ideal taper means the phase velocity is constant along the taper while the non-uniform $\epsilon_{r}[n]$ of the PUT results in a non-uniform phase velocity, $v_p[n]$. (\ref{thetaeqn}) can also be expressed in terms of layer thickness $l_N$ as:

\begin{equation} \label{lN}
  l_{N} = \frac{\sqrt{\REV{\epsilon_{eff}}}L}{\sum_{n=1}^{N}\sqrt{\epsilon_{r}[n]}},
\end{equation}

\noindent \REV{which is a constant $l_{N}=L/N$ for the effective permittivity which satisfies $\sqrt{\epsilon_{eff}}=\frac{1}{N}\sum_{n=1}^{N}\sqrt{\epsilon_{r}[n]}$}. This condition results in the electrical length and physical length of the PUT being equal to that of the ideal taper. Therefore the transform starts from phase equality and\REV{, with the proper $\epsilon_{eff}$,} ends up with uniform layer thickness equality. 

\begin{algorithm}[t]
\begin{minipage}{\columnwidth}

	\caption{Non-commensurate Taper Design} 
	\label{alg:taper}
	\begin{algorithmic}[1]
	\renewcommand{\thefootnote}{\fnsymbol{footnote}}
		\REQUIRE $Z_{01}$, $Z_{02}$, $L$, $N$, $f_{c}$\footnotemark[4], $TOL$
		\ENSURE $ Z[n], \epsilon_r[n]$, $\epsilon_{eff}$
		\STATE calculate $\epsilon_{r01}$ and $\epsilon_{r02}$ 
		\STATE set $ \epsilon_{eff,L} = \epsilon_{r01}$, $ \epsilon_{eff,R} = \epsilon_{r02}$,
		
		$ \epsilon_{eff,M} = \frac{ \epsilon_{eff,L}+ \epsilon_{eff,R}}{2}$, $e >TOL$
		\WHILE{$e>TOL$ }
		\STATE  $A_{L/R/M}$ using (\ref{ga and band})\footnotemark[4]
		\STATE  $Z(z)_{L/R/M}$ using \cite{Pozar05Mwave}
		\STATE  $Z[n]_{L/R/M}$ for PUTs
		\STATE  $\epsilon_{r}[n]_{L/R/M}$ for PUTs
		\STATE  $l_{N,L/R/M}$ using (\ref{lN}) 
		\STATE  errors, $e_{L/R/M} = \frac{L}{N} - l_{N,L/R/M}$   
		\IF{$e_L$ and $e_R$ have opposite signs}
        \STATE$ \epsilon_{eff,R} =  \epsilon_{eff,M}$,{ exact solution is in this interval}
        \ELSE
        \STATE$ \epsilon_{eff,L} =  \epsilon_{eff,M}$,{ exact solution is in this interval}
        \ENDIF
        \STATE$ \epsilon_{eff,M} = \frac{ \epsilon_{eff,L}+ \epsilon_{eff,R}}{2}$,
        \STATE $e$ = min\{$e_{L/R/M}$\},
        \STATE choose $ \epsilon_{eff}$,  $\epsilon_{r}[n]$ and $Z[n]$ with the smallest $e$
        \ENDWHILE
		\RETURN $Z[n], \epsilon_r[n]$, $\epsilon_{eff}$ 
		\vspace{-0.25cm}
		\footnotetext{$\S$These are only for Klopfenstein design.}

	\end{algorithmic}
\end{minipage}

\end{algorithm}

However, (\ref{lN}) is a transcendental function (because $\epsilon_r[n]$ must be derived for the Klopfenstein profile) which cannot be directly solved to find $\epsilon_{eff}$. Therefore Algorithm\,\ref{alg:taper} presents an iterative bisection method for the design of PUTs, which also facilitates the design of other types of tapers. The inputs are the desired port impedances, $Z_{01}$ and $Z_{02}$ , taper length $L$, number of layers $N$, and taper cutoff frequency $f_{c}$. Since the bisection method is iterative, an error tolerance, $TOL$, is set for convergence and is specific to different taper parameters. The proper ${\epsilon_{eff}}$ and corresponding $ \epsilon_r[n]$ are found through iteration---to realize bisection three ideal tapers are calculated in each iteration: ${\epsilon_{eff,L}} ={\epsilon_{r01}} $ is the left (or lower) boundary, ${\epsilon_{eff,R} ={\epsilon_{r02}}}$ is the right (or upper) boundary, and ${\epsilon_{eff,M} = \frac{\epsilon_{eff,L}+\epsilon_{eff,R}}{2}}$ is the mean. Then three PUTs are designed with three layer thicknesses, $l_{N,L/R/M}$. A suitable solution is reached when the error between $l_{N,L/R/M}$ and $\frac{L}{N}$ is less than $TOL$ so in each iteration of the algorithm a length error is calculated for the three cases as $e_{L/R/M} = \frac{L}{N} - l_{N,L/R/M}$. Then new boundary permittivity values are selected for the next loop and the PUT permittivity associated with the current loop is assigned that profile with the smallest error. For the bisection method, errors $e_{L/R/M}$ with different signs enclose the exact solution so the new range is selected to include the exact solution. Precise calculations are presented in Algorithm\,\ref{alg:taper}. Note that except $f_c$ and Step 4 (unique to Klopfenstein), the discretization and transformation methods are generally applicable for any taper if the corresponding impedance profile is chosen in Step 5. \REV{Once $\epsilon_{eff}$ is found to equalize electrical and physical length, the frequency response of the ideal Klopfenstein taper for a non-commensurate line can be found as:}

    \begin{multline}
    \label{gammaKlop}
    \REV{\left|\Gamma(f)\right|=}\REV{\left|\Gamma_0\exp(-j\beta_{eff}L)\frac{\cos\sqrt{(\beta_{eff}L)^2-A^2}}{\cosh{A}}\right|}\\
    \REV{=\left|\Gamma_0\exp(-j\frac{2\pi f\sqrt{\epsilon_{eff}}}{c_0}L)\frac{\cos\sqrt{(\frac{2\pi f\sqrt{\epsilon_{eff}}}{c_0}L)^2-A^2}}{\cosh{A}}\right|.}
    \end{multline}

\noindent \REV{$\beta_{eff}$ is the effective propagation coefficient and depends upon $\epsilon_{eff}$. This expression is similar to the classical expression for the frequency response of a Klopfenstein taper except that it has been written in terms of the proper $\epsilon_{eff}$ and can now predict the correct cutoff frequency and passband performance.}

\REV{In the following sections three different Klopfenstein tapers will be discussed so we wish to clarify the purpose behind the three different tapers: i) The first taper, discussed in Section\,\ref{sec:impairment}, is a nine layer taper matching $Z_{01}=308\Omega$ to $Z_{02}=184\Omega$---this taper was chosen because it clearly shows the impairments we wish to discuss. ii) The second taper is the fabricated taper, discussed in Section\,\ref{sec:prototype}, and also comprises nine layers but matches a more extreme range from $Z_{01}=308\Omega$ to $Z_{02}=140\Omega$---this is representative of a maximum core permittivity in our typical fabricated lenses. iii) The third taper is a six layer taper used in the simulated Klopfenstein-matched demonstration lenses in Section\,\ref{sec:lens}---this taper only uses six layers in order to reduce the complexity (and also cost) of the matched GRIN lenses to make them as realistic as possible. In practice we try to minimize the number of layers in a matching section because cost is dominated by the layers in the taper (the core can be fabricated from a small number of thick substrates).} \SREV{(in this manuscript we study a unit-cell which could be used in [1]).}

\section{Nonidealities of PUTs} \label{sec:impairment}

\begin{figure}[t]
    \centering
    \includegraphics[width=0.47\textwidth]{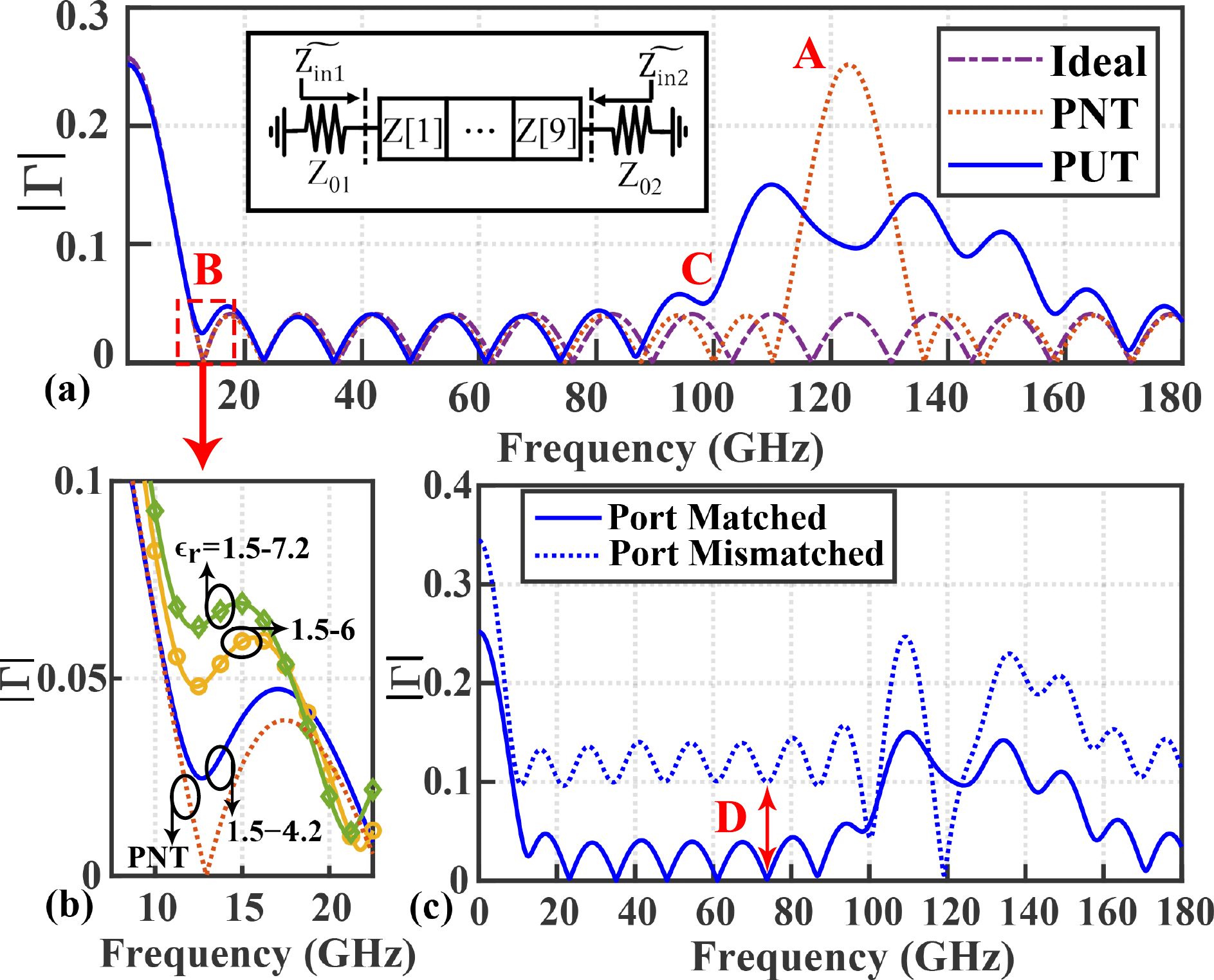}
    \caption{Frequency response of the proposed taper. (a) Ideal, PNT, and PUT with $Z_{01}$=308$\Omega$ and  $Z_{02}$=184$\Omega$. (b) First zero shift of PUT caused by different relative permittivity matching ranges. (c) Comparison among the original PUT with $Z_{01}$=308$\Omega$, $\widetilde{Z_{in1}}$=308$\Omega$ and $Z[1]$=289$\Omega$; method (i) with $Z_{01}$=377$\Omega$, $\widetilde{Z_{in1}}$=308$\Omega$ at $f_0$ and $Z[1]$=289$\Omega$; and method (ii) with $Z_{01}$=377$\Omega$, $\widetilde{Z_{in1}}$=377$\Omega$ at $f_0$, $Z[1]$=289$\Omega$.} 
    \label{fig:twofig}
\vspace{-0.25cm}
\end{figure}

\REV{As mentioned, }the case study PUT uses nine layers to provide an impedance match from the realizable permittivity values of $\epsilon_r$\,=\,$1.5$ ($Z_{01}$\,=\,$308\Omega$) to a core permittivity of $\epsilon_{rc}$\,=\,$4.2$ ($Z_{02}$\,=\,$184\Omega$), \REV{which is instructive to show the taper properties as stated above. The schematic is shown in the inset of Fig.\,\ref{fig:twofig}(a) where each of the nine substrate layers is $l_N=0.030"$ (total thickness is $0.27"$) and $f_{c}$\,=\,$11$\,GHz}. As mentioned above, the idealized impedance profile was discretized by selecting the central impedance of $N$\,=\,$9$ sections from the ideal taper. Algorithm\,\ref{alg:taper} produced the following outputs: $\epsilon_{eff}$\,=\,$2.56$, $Z[n]$\,=\,290, 279, 266, 252, 238, 224, 212, 202, 195\,$(\Omega)$, $\epsilon_r[n]$\,=\,1.68, 1.82, 2.00, 2.23, 2.51, 2.82, 3.15, 3.47, 3.75, and $\Gamma_{max}$\,=\,$0.04$. \REV{These results satisfy the electrical and physical length equality condition in (\ref{lN}).} At $f_0$ the Klopfenstein taper achieves a perfect impedance match where $\widetilde{Z_{in1}}$\,=\,$Z_{01}$ (see inset of Fig.\,\ref{fig:twofig}(a)) and $\widetilde{Z_{in2}}$\,=\,$Z_{02}$. Figure\,\ref{fig:twofig}(a) shows the response of the ideal taper (calculated \SREV{as $\Gamma(\theta)$ from [13]}\REV{with (\ref{gammaKlop}) and using $\epsilon_{eff}$}), the PNT (\REV{using (\ref{cf})}) and the PUT (calculated with wave-transfer matrices \cite{saleh2019fundamentals}). There are a number of nonidealities present in both the PNT and PUT which are indicated with letters ``A'' to ``D'' in Fig.\,\ref{fig:twofig}(a) and (c). The remainder of this section discusses each nonideality in detail and provides suggestions for mitigating each, if possible.

\textbf{``A'':} 
The continuous Klopfenstein taper is a high-pass structure while the discretized PNT is approximately a Chebyshev transformer with a finite passband resulting from discretization with finite-thickness layers. From (\ref{cf}), each layer of the PNT is a quarter-wavelength transformer at $f_0=61.6\,$GHz. Therefore, at $2f_0$ each layer will be a half-wavelength long and the taper comprising $N$ layers will be $N\lambda/2$ in length. The taper then mirrors the impedance of the second port at the first resulting in $\Gamma$ at ``A'' being equal to $\Gamma_0$.
    
\textbf{``B'' and ``C'':} These impairments are similar to ``A'' but for a non-commensurate taper (the PUT) in which multiple reflections add incoherently over a range of frequencies around $2f_0$. For a PNT the center frequency $f_0$ at which each layer is a quarter-wavelength long (from (\ref{cf})) is 61.6\,GHz while for a PUT $f_0$ and $\theta$ (at 62.1\,GHz) of each layer varies as shown in Table\,\ref{table}. For the fixed layer thickness of the PUT and defining the center-most layer (with $\epsilon_r=2.51$) to be a quarter wavelength at $62.1\,$GHz, the corresponding $\theta$ of the outermost layers ranges from $110^\circ$ to $74^\circ$ (in other words, they are non-commensurate). Layers in the center of the PUT are approximately 90$^\circ$, so that the central part of the passband response is consistent with the PNT. Toward the edges layers deviate more significantly from 90$^\circ$ resulting in an increase in $\Gamma$ and distortion at both the low frequency band edge (``B'') and the high frequency band edge (``C''). The distortion at ``C'' was also observed in \cite{Mrakami98TD,Lee11SIW}. At ``C'' there is a smearing of the band edge compared to ``A'' because there is a range of frequencies at which various layers are a half-wavelength. This smeared high-frequency cutoff approximately spans from $2\times50.8$\,GHz ($101.6$\,GHz) to $2\times75.9$\,GHz ($151.8$\,GHz). The result of these incoherent reflections is a lower reflection (compared to $\Gamma_0$ at ``A'') but a significant reduction in the high frequency band edge which must be considered in a design. The impedance transformation ratio influences the central frequencies of each layer, thus influencing ``B'' and ``C''. A smaller impedance transformation ratio minimize this effect (see zoom of Fig.\,\ref{fig:twofig}(b)).

\begin{table}[t]
\begin{minipage}{\columnwidth}
\centering
\caption{Layer profile of a PUT$^\dagger$}
\renewcommand{\arraystretch}{1.2}
\setlength{\tabcolsep}{3pt}
\begin{tabular}{@{}rcccccccccc@{}} 
\toprule
     Layer       & 1    & 2    & 3    & 4    & 5    & 6    & 7    & 8    & 9    \\ 
\midrule
$f_0$ (GHz) & 50.8 & 52.9 & 53.7 & 56.7 & 62.1 & 65.9 & 69.6 & 73.0 & 75.9 \\ 
    $\theta^\ddagger (\deg)$ & 110.0 & 105.8 & 104.2  & 98.6 & 90.0 & 84.9 & 80.3 & 76.6 & 73.6\\ 
\bottomrule
\end{tabular}
\footnotetext{$\dagger$For a PNT, each layer has an identical $f_0=61.6$\,GHz and $\theta=90^\circ$.}
\footnotetext{$\ddagger$Phase at central frequency $f_0=62.1$\,GHz.}
\vspace{-0.25cm}
\label{table}
\end{minipage}
\vspace{-0.25cm}
\end{table}

\textbf{``D'':} As mentioned above, a perfect impedance match implies $\widetilde{Z_{in1}} = Z_{01}$ at $f_0$. However, since most artificial dielectrics have a lower permittivity limit (suppose $\epsilon_{r,min}=1.7$), there is an upper limit on the realizable impedance within the taper (correspondingly, $Z_{max}=289\Omega$). And yet lenses typically interface with waves in free-space, $Z_{01}=377\Omega$. There are two approaches to mitigating this issue, shown in Fig.\,\ref{fig:twofig}(c): Method (i) is to design the match to the realizable $Z_{max}$ and then accrue an additional mismatch at the outer surface relative to free-space causing $Z[1]=Z_{max}$ and $Z[1]<{\widetilde{Z_{in1}}}<Z_{01}$, or specifically for a Klopfenstein taper method (ii) is to use (\ref{ga and band}) to design a taper with increased $\Gamma_{max}$ such that the edge discontinuity expects a mismatch to free-space where $Z[1]=Z_{max}$ and $Z[1]<{\widetilde{Z_{in1}}}=Z_{01}$. While method (ii) seems compelling because it advantageously uses the impedance discontinuity at the edges of a Klopfenstein taper, Fig.\,\ref{fig:twofig}(c) shows that it is much worse than method (i). The increase of $\Gamma$ by mismatch is labeled as ``D" and estimated as $\Delta\Gamma=\left(Z_{01}-\widetilde{Z_{in1}}\right)/\left(Z_{01}+\widetilde{Z_{in1}}\right)$ in Fig.\,\ref{fig:twofig}(c).



\section{Taper Prototype} \label{sec:prototype}

\begin{figure}[t]
    \centering
    \includegraphics[width=\columnwidth]{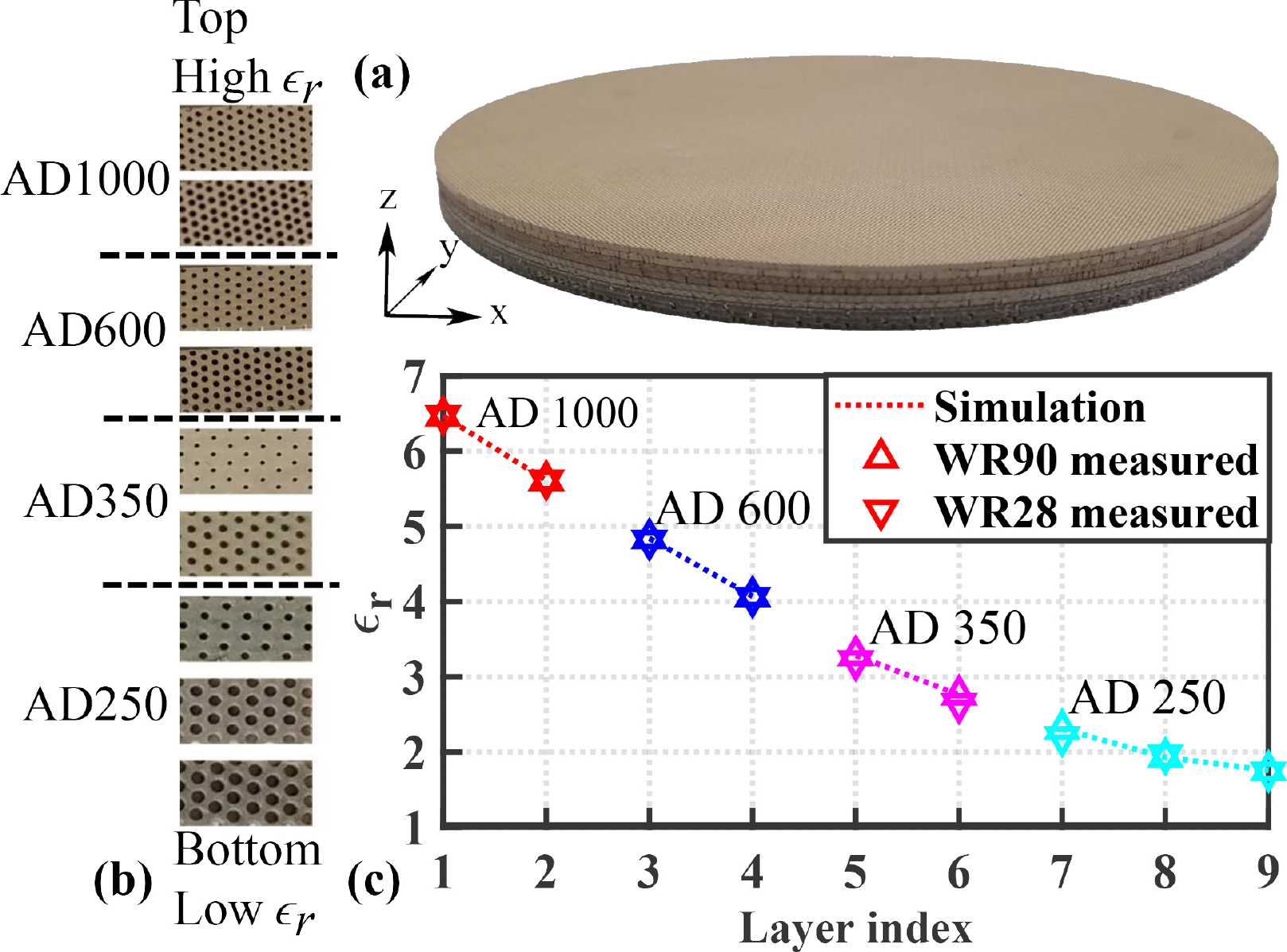}
    \caption{(a) Fully assembled 4-inch taper. (b) Zoom of each layer of perforated dielectric. (c) Measured and simulated effective permittivity from free-space characterization.}
    \label{fig:material}
\end{figure}

To demonstrate the proposed method we designed and fabricated a prototype Klopfenstein PUT with $N=9$ layers, each of $l_N=0.030"$ (total thickness 0.27"). This taper design differs from the case study considered in the previous section---the case study was designed to clearly demonstrate the various impairments (and thus it matched to a moderate core permittivity of 4.2) while this prototype matches to a high core relative permittivity of 7.2 in order to validate the performance of one of the most extreme unit-cells in a hypothetical GRIN lens. Referring to Fig.\,\ref{fig:structure}(a), the unit-cell corresponds to the black dashed boxes and the prototype taper is the impedance matching layers along the $z-$axis from free-space at the bottom of the GRIN lens to $\epsilon_{rc}=7.2$ ($Z_{02}=140\Omega$) in the core of the GRIN lens. Since the core permittivity of a GRIN lens varies radially the required taper across the bottom (and top) of the lens must also vary radially. Here we only fabricate a single instance of a 0.27" taper with permittivity variation along the $z-$axis and uniform permittivity in the cross section ($xy-$plane). The cross section diameter is 4" to ensure the majority of the energy from our free-space measurement setup (described below) intercepts a uniform taper cross section.

\begin{figure}[t]
    \centering
    \includegraphics[width=0.85\columnwidth]{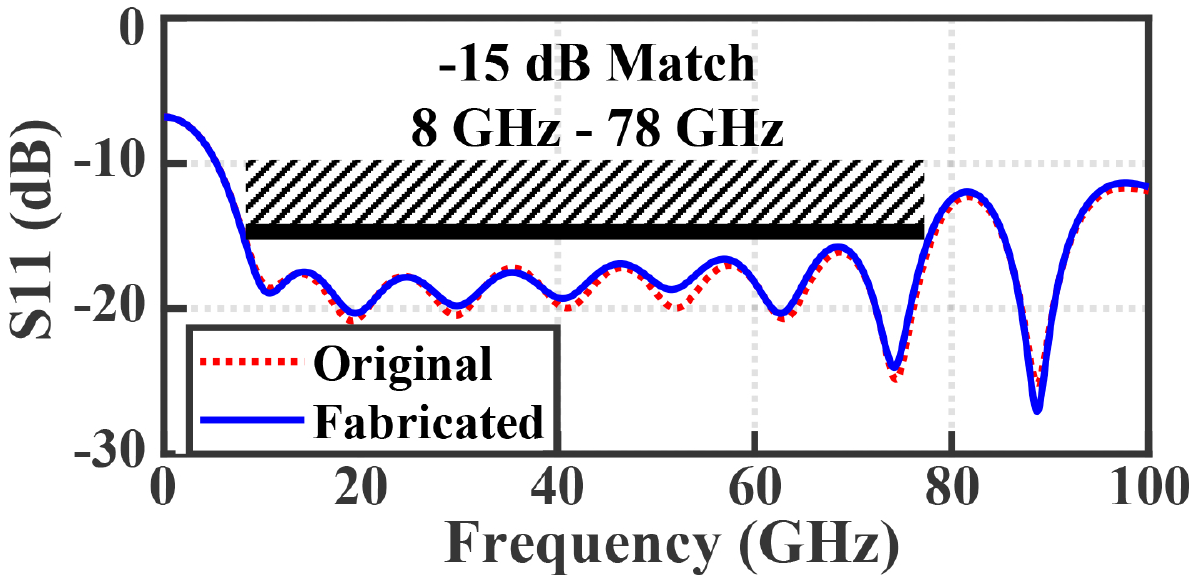}
    \caption{Taper response of original and fabricated tapers with \mbox{-15\,dB} match bandwidth indicated from $8$--$78$\,GHz.}
    \label{fig:resp}
    \vspace{-0.2cm}
\end{figure}

The original taper was designed using Algorithm\,\ref{alg:taper} to match with $Z_{01}=308\Omega$ and $Z_{02}=140\Omega$ (method (i) in Fig.\,\ref{fig:twofig}(c)) with an $f_c=11$\,GHz. The resulting permittivity profile is $\epsilon_{r}[n]=1.70, 1.92, 2.24, 2.69, 3.29, 4.01, 4.82, 5.63, 6.34$ and $\epsilon_{ri}=3.46$ \REV{(these results also satisfy (\ref{lN}))}. The realized taper is shown in Fig. \,\ref{fig:material}(a) where all nine layers have been stacked and bonded. Rectangular samples of each layer are shown in Fig.\,\ref{fig:material}(b). The effective permittivity of each layer was realized with a perforated dielectric as described in \cite{Garcia_Matched_2020,garcia_silicon_2017} and results in a slight modification to the original permittivity values. To span the \REV{modified} permittivity range 6.43--1.74, holes were drilled on a hexagonal lattice in four substrates (Rogers AD1000, AD600, AD350, and AD250) with nominal permittivity values of 10.35, 6.15, 3.5 and 2.5, respectively. Drill diameters were 10.5, 11.8\,mil in AD1000 (18\,mil lattice constant), 10.5, 13.8\,mil in AD600 (23\,mil lattice constant), 8.3, 16\,mil in AD350 (32\,mil lattice constant), and 13.8, 21.7, and 25.6\,mil in AD250 (37\,mil lattice constant). The frequency response of the originally designed permittivity profile and the modified profile considering fabrication (``Original'' and ``Fabricated'') are shown in Fig.\,\ref{fig:resp} with very little deviation. The passband of a 15\,dB return loss, is 8-78\,GHz. 

\begin{figure}[t]
    \centering
    \includegraphics[width=0.9\columnwidth]{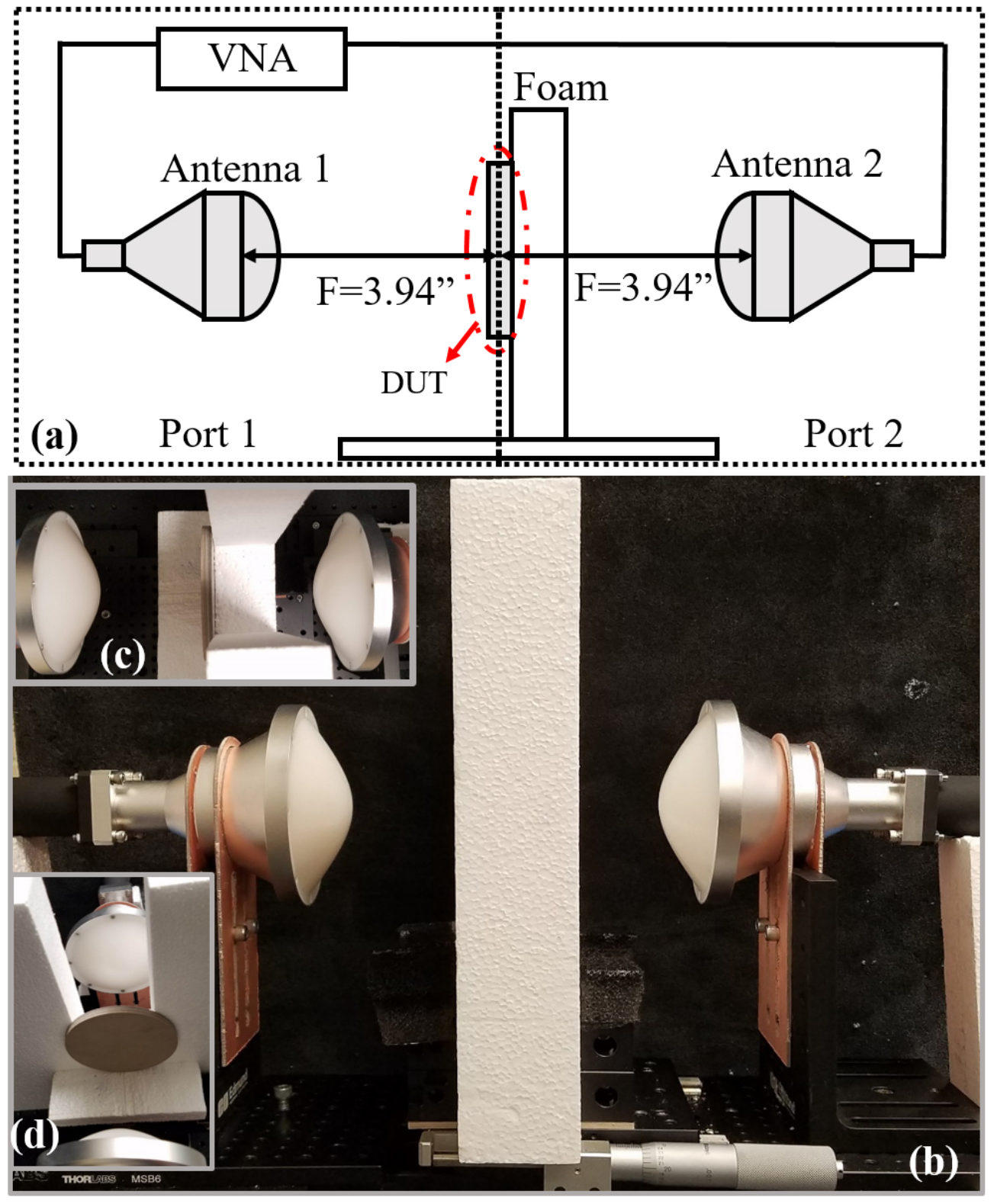}
    \caption{Measurement Setup in the WR90 band: (a) Schematic of free-space measurement setup. (b) Side-view without DUT. (c) Top-view with DUT. (d) Front-view with DUT.}
    \label{fig:setup}
    \vspace{-0.25cm}
\end{figure}

The 4" diameter disk of each separate layer was measured in a free-space material characterization setup, shown in Fig.\,\ref{fig:setup}. Since the tapers are intended for use in quasi-optical structures and lens antennas it is important that the measurement setup closely mimic the intended use case with nearly TEM incident radiation. This prevented us from measuring the tapers in a rectangular waveguide as was done in \cite{Garcia_Matched_2020,garcia_silicon_2017}. Instead, the free-space setup employs spot-focusing antennas in the WR90, WR28 and WR12 waveguide bands to produce approximately collimated TEM waves within a confined area. Specifically, 83\% of the energy from the spot antennas is contained within a 2.51", 1.2" and 0.28" diameter, respectively. \REV{To validate the TEM assumption in the measurement, transmission efficiency versus incident angle of the fabricated taper was simulated for both TE- and TM-modes. At normal incidence transmission efficiency is 98.7\% for both modes, and 99.8\%(95.3\%) for TM(TE) mode at 45$^\circ$ (note that the TM-mode is approaching Brewster's angle and thus transmission increases off-broadside). Since spot-focusing antennas are specifically designed to achieve a nearly collimated wavefront at their focal distance $F_{spot}$, the wave will have close to normal incidence at the samples. However, to be conservative we calculate the largest possible incident angle from a spot-focusing antenna of diameter $D_{spot}$ as $\arctan(D_{spot}/2F_{spot})$. For the X, Ka and W-band antennas used in the measurements the angle is 30$^\circ$, 21$^\circ$ and 30$^\circ$, respectively with transmission  efficiency greater than 99.1\%(97.7\%) for TM(TE) mode. Therefore the TEM assumption is valid in the free space measurement setup.}

Each layer of the taper was individually mounted to one side of the low dielectric mounting foam shown in Fig.\,\ref{fig:setup}. A 3" slot was removed from the middle of the foam to further decrease the influence of the foam. A two tier calibration was used: first an SOLT calibration was used to deembed systematic errors in the VNA and the coaxial cable, then the Gate-Reflect-Line (GRL) calibration method \cite{BegleyGRL2012} was used to extend the phase reference plane from the coaxial cable to the surface of the taper and decrease multipath reflections. The measured permittivity values of each layer, extracted at 10.3\,GHz in the WR90 band and 33.3\,GHz in the WR28 band, are shown in Fig.\,\ref{fig:material}(c) along with the simulated permittivity profile and they show very close agreement in both waveguide bands. This measurement validates that each layer was fabricated correctly and that the measurement setup and calibration method were effective. 

\begin{figure}[t]
    \centering
    \includegraphics[width=0.85\columnwidth]{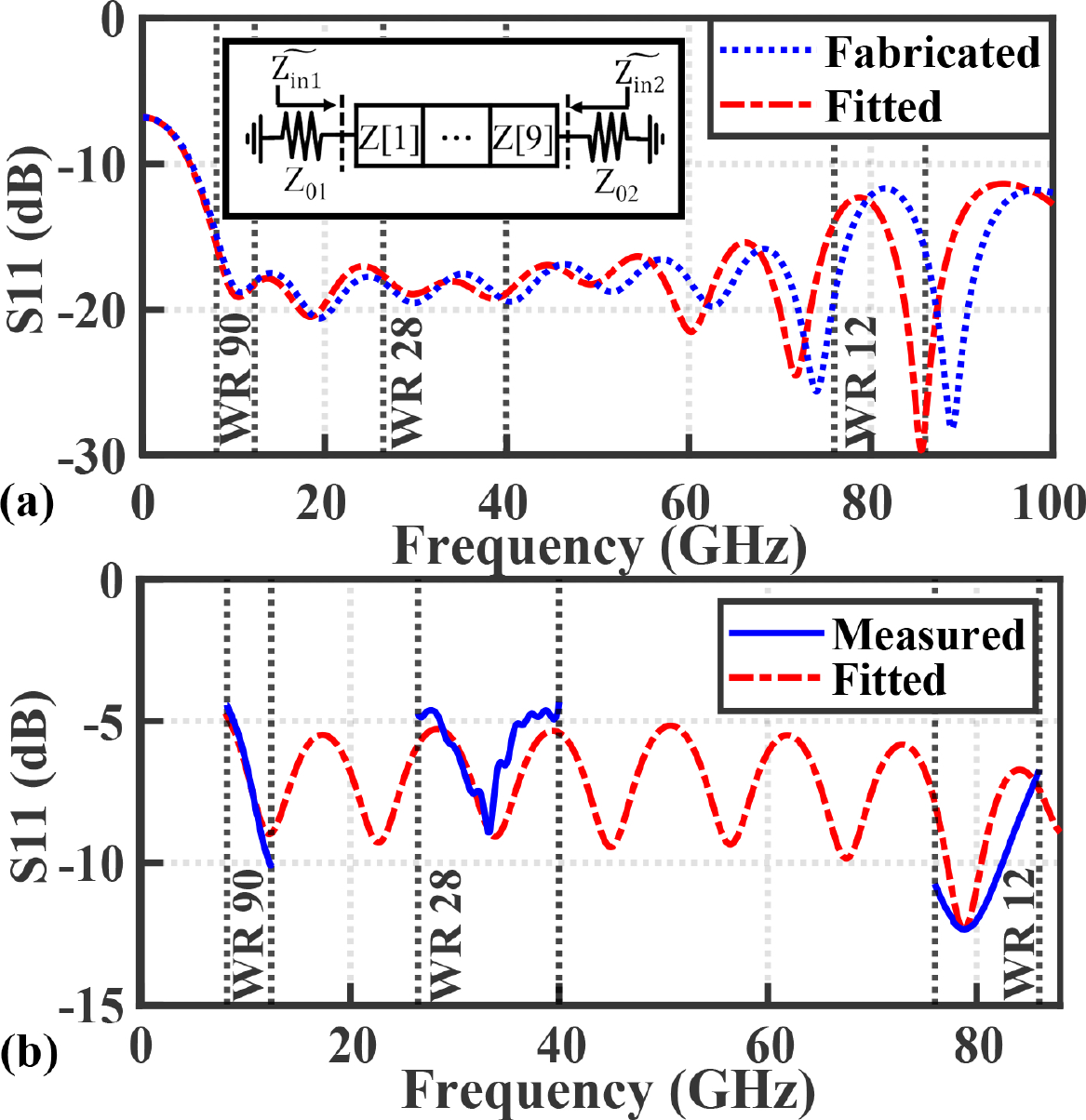}
    \caption{(a) Comparison between the fabricated  taper and fitted taper.(b) Measured taper response and renormalized fitted response from free-space spot-antenna measurements in the WR90, WR28 and WR12 waveguide bands. }
    \label{fig:final}
    \vspace{-0.25cm}
\end{figure}

After confirming accurate fabrication of each layer the complete taper was bonded with a total thickness of 284\,mil (12\,mil thicker than nominal). The increase in thickness causes a decrease in the cutoff frequency and each null of the passband response. Fig. \,\ref{fig:final}(a) shows the simulated response of the taper using the predicted permittivity values (``Fabricated'', same as blue curve in Fig. \ref{fig:resp}) and a fitted taper (``Fitted'') which uses the average measured permittivity values reported in Fig.\,\ref{fig:material}(c) and slightly thicker PCB layers to account for manufacturing tolerances (discussed in detail below). 

The prototype taper was measured with the same free-space measurement setup as described above. The simulated taper match bandwidth spans at least seven waveguide bands (from WR137 to WR12). Due to setup limitations the match response was measured in three bands: near taper cutoff, $f_c$, in the WR90 band (8.2--12.4\,GHz), in the taper pass-band in the WR28 band (26.5--40\,GHz) and at the taper high-frequency cutoff in the WR12 band (76--86\,GHz, limited by the spot antenna). The measured results are shown in Fig.\,\ref{fig:final}(b). Since the second port impedance was designed to be $Z_{02}=140\Omega$, but was measured in free-space ($Z_{02}=377\Omega$), we renormalized the simulated taper response \cite{Frickeyrenormn1994} 
to confirm the measurement. In order to achieve good agreement between the measured prototype and simulation (``Fabricated''), Monte-Carlo simulations with layer thicknesses ranging over the expected tolerances were conducted and the fitted response  in Fig.\,\ref{fig:final}(b) corresponds to a taper with the following substrate thicknesses: 31.6\,mil for AD1000, 30.8\,mil for AD600, 31.6\,mil for AD350, and 31.4\,mil for AD250. These thicknesses increase the overall thickness of the taper by 10.2\,mil, close to the measured thickness increase of 12\,mil. 
Fig. \,\ref{fig:final}(a) shows the nominal design (``Fabricated") and the fitted design with good agreement, confirming both the design and the fabrication. The results of this measurement confirm the validity of the design method presented in Algorithm\,\ref{alg:taper}.

\section{Demonstration Lens Design \& Simulation} \label{sec:lens}

In this section we demonstrate the effectiveness of the proposed design method with simulations of several complete GRIN lens designs. \REV{First the GRIN lens modeling and simulation setup are introduced, then we present an approximate expression for the efficiency of a GRIN lens which serves as a design guide. Finally, several Klopfenstein taper GRIN lenses are designed and their performance is compared with an unmatched lens, as well as GRIN lenses matched with quarter-wave sections and exponential tapers.}

\subsection{\REV{Lens Design and Simulation Setup}}

\REV{All of the lenses designed in this section follow the procedure outlined in \cite{Garcia_Matched_2020} and summarized as follows: a GRIN lens comprises 250 concentric rings arranged radially from the center of the lens across a diameter of 8" ($2.7\lambda_0$ at 4\,GHz). Each ring is a unit-cell comprising a matching layer on the bottom and top of a uniform core permittivity and has an overall thickness of 1.2". The unit-cells are arranged such that their respective absolute phase delays equalize the optical path length from a central feed element (at a focal distance of 4" from the bottom surface of the lens) to the top surface of the lens to produce a collimated wave and far-field beam at broadside. The unit-cells are selected from a library of unit-cells which include a six layer matching taper on the top and bottom of a uniform core with permittivity values ranging from $\epsilon_{rmin}$ to $\epsilon_{rmax}$ (corresponding to e.g., the perforated dielectrics as described in \cite{Garcia_Matched_2020}). The focal length and diameter of the GRIN lens ($F/D=0.5$) determine the amount of phase which must be collimated. The thickness of a unit-cell is determined by the range of available permittivities and the amount of phase to be collimated. In order to demonstrate the influence of the taper design on GRIN lens performance, several lenses with different matching sections are presented in the next two subsections. We chose to use six 30-mil layers in the tapers (both for Klopfenstein and Exponential taper matched lenses) because it clearly reveals the taper properties and reduces overall complexity of the lens. In order to maintain clear and predictable trends in the simulated lens responses $\epsilon_{rmin}=1$. In practice a higher minimum permittivity would result in slightly higher reflections.} 

\REV{The designed lenses were simulated with high fidelity over a very wide frequency range using an in-house two-dimensional (2D) finite-difference time domain (FDTD) code with an idealized feed. Two-dimensional simulations were used instead of three-dimensional in order to manage the computational complexity of the GRIN lenses over such large frequency spans. It is supposed that in the simulation the $z$-plane (normal to the 2D simulation plane) is infinite and there is no $z$ variation in the fields---this 2D assumption does not sacrifice the generality of the conclusions drawn in this section since most lenses are designed with azimuthal symmetry. The simulation frequency is from 4\,GHz to 60\,GHz and the grid was chosen to be $\lambda_0/50$ at 60\,GHz ($\Delta X=0.1$\,mm) in order to provide for very accurate aperture fields. This was necessary to be able to differentiate between several lenses which all have extremely high efficiency. The TE-mode was simulated to obtain the worst-case transmission result (the TM-mode has a transmission peak at Brewster's angle while the TE-mode does not). An ideal 2D isotropic point source excitation was used so that the illumination efficiency remained constant over frequency and all variation in efficiency over frequency could be attributed to the lens and matching sections. The far-field radiation pattern was computed from the aperture fields along a line of extent $D=8$" and compared to the gain from an ideal uniform line source, $2D/\lambda$. For a fair comparison of different matching approaches, the efficiency of each lens should be dominated by matching performance and not, e.g., poor phase collimation. Therefore, each lens was designed with an iterative method described in more detail in Section\,\ref{sec:lens:anal} to achieve nearly perfect phase collimation (relative to a wavelength at 60\,GHz).}
    
\subsection{\REV{Performance Estimation and Full-wave Simulation}}   \label{sec:lens:anal}

   \begin{figure}[t]
    \centering
    \includegraphics[width=8.2cm]{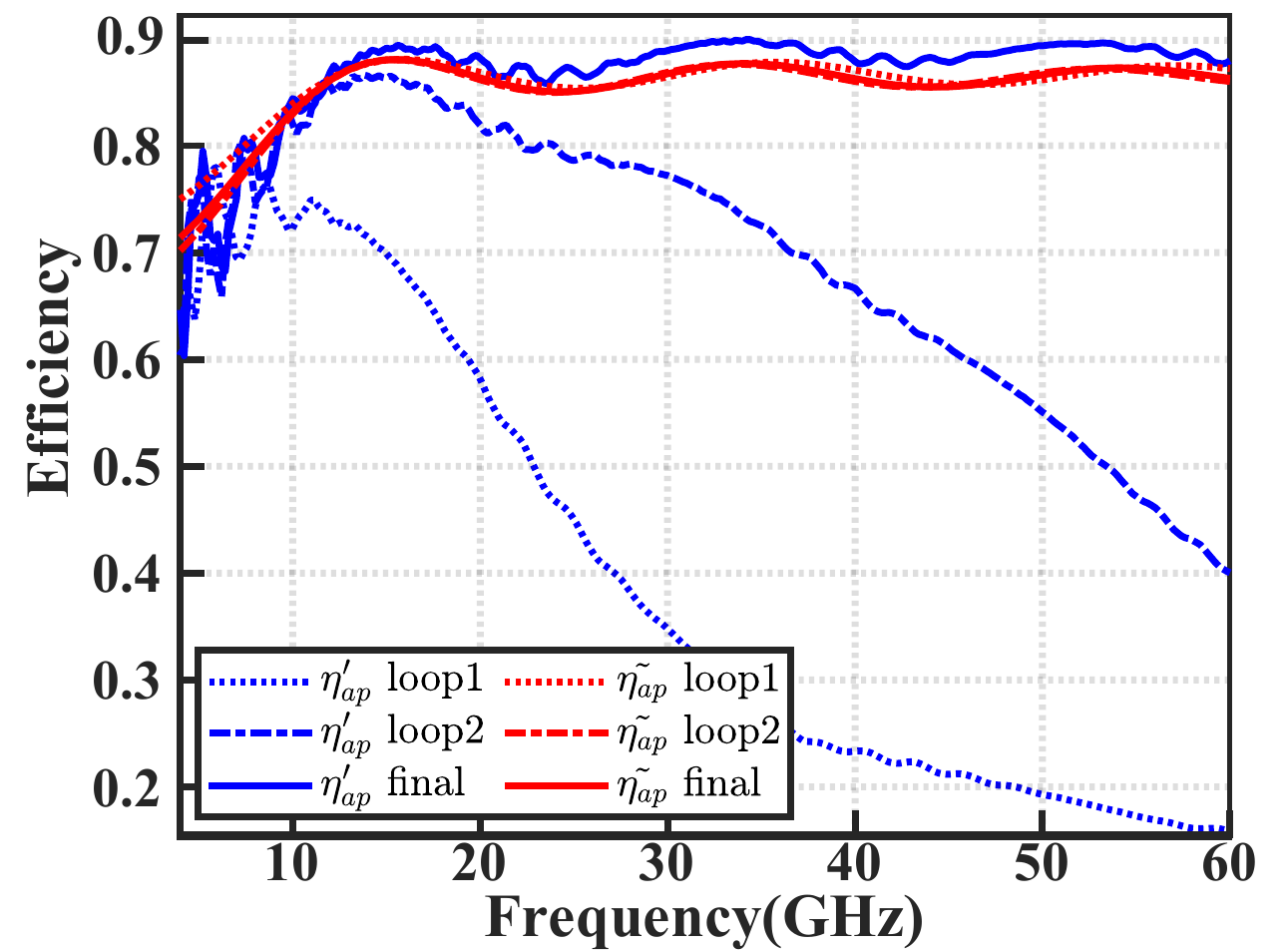}
    \caption{\REV{Iterative design of a Klopfenstein taper GRIN lens. Taper are six layers per side with $f_c=11$\,GHz. Blue and red traces are $\eta_{ap}'$ (see (\ref{eq:ApPrime})) and $\widetilde{\eta_{ap}}$ (see (\ref{ap_etaT})).}}
    \label{fig:iteration loop}
    \vspace{-0.25cm}
    \end{figure}

\REV{The performance of the designed lenses can be evaluated with a full-wave simulation code (e.g., the 2D-FDTD code described above or a 3D code). But, even an efficient 2D code can take hours to complete due to the complexity of GRIN lenses (our 2D-FDTD code with the fine simulation grid mentioned requires approximately 4 hours to solve from 4\,GHz to 60\,GHz) and, due to approximations made in the design algorithms, the result will be a GRIN lens with imperfect phase collimation. The challenge is that since GRIN lenses provide so many degrees of freedom, analytical design expressions are typically derived from simplified straight-line ray tracing \cite{MaCui3_APL_2011,isakov_3d-printed_2016} or curved ray tracing without matching layers \cite{Zhang_curved_raytracing_2020}. Due to the approximations in these methods phase collimation is imperfect and therefore gain suffers over very wide operating bandwidths. In these cases iterative or optimization methods are used to modify the initial lens design. Suppose the GRIN lens materials are non-dispersive: then if the phase distribution is derived and modified (through iteration) at the highest frequency of interest, the lower frequency band will also exhibit good phase collimation. Thus a general way to modify the wideband design is to record the phase distribution error at the design frequency in the current iteration and then modify the phase correction by the current error until the efficiency converges.} 

\REV{Figure \ref{fig:iteration loop} shows lens efficiency (defined in (\ref{eq:ApPrime}) and (\ref{ap_etaT}), below) for the first, second, and final (6th) iteration of the design method in \cite{Garcia_Matched_2020}. The blue traces show $\eta_{ap}'$ (see (\ref{eq:ApPrime}), below) improving over a wide bandwidth because of increasingly better phase collimation throughout the iteration. However, it is time consuming to evaluate different taper designs in order to identify one that meets the design specification. For this reason we seek an approximate expression for aperture efficiency which does not require a full-wave simulation and can serve as a target for iterative designs. Specifically, based on the initial lens profile (first iteration), a formula for approximate aperture efficiency, $\widetilde{\eta_{ap}}$ (red traces in Fig.\,\ref{fig:iteration loop}) can be derived which then serves as a target for iteration---convergence is reached when the simulated lens is close enough to the target. The approximate aperture efficiency also guides the initial specification of the unit-cell (e.g., total thickness, number of discrete layers in the tapers, required minimum and maximum permittivity) to achieve a desired passband efficiency and low-frequency cutoff.}

   \begin{figure*}[t]
    \centering
    \includegraphics[width=19cm]{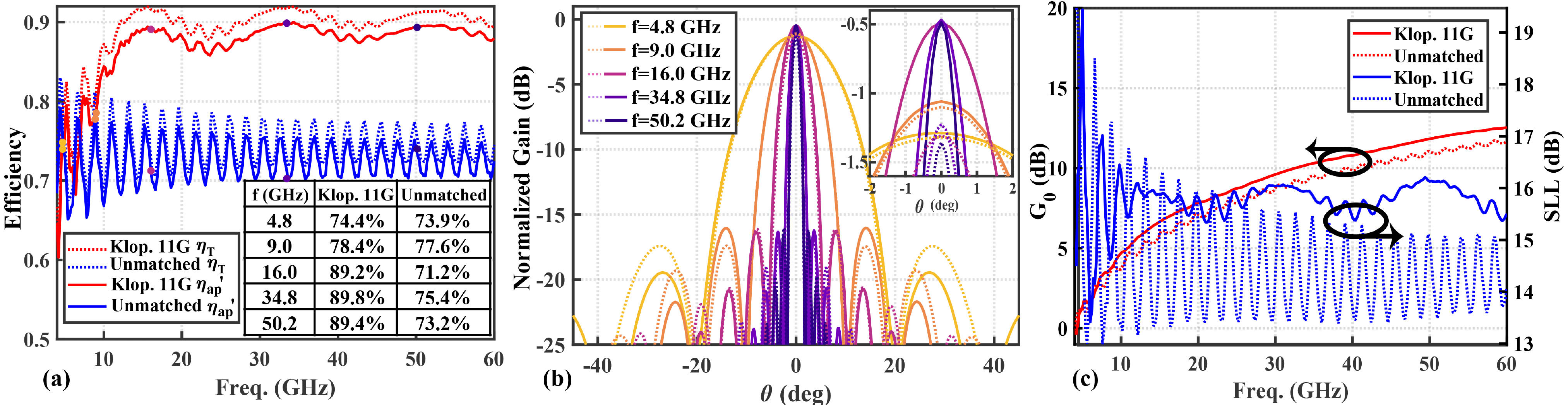}
    \caption{\REVNEW{(a) Transmission efficiency $\eta_T$ and normalized aperture efficiency $\eta_{ap}'$ versus frequency of a  Klopfenstein taper matched lens and an unmatched lens. The inset table shows the marked $\eta_{ap}'$ values.  (b) Gain patterns of the Klopfenstein matched lens (solid) versus the  unmatched lens (dotted), which are normalized to $\eta_{ap}'=1$ at each frequency. The inset figure showsshows a zoom view of the pattern peaks where the vertical axis of the inset is Normalized gain in dB. (c) Gain maxima $G_o$ and sidelobe level of the Klopfenstein-matched lens versus the unmatched lens. }}
    \label{fig:matVSunmat}
    \vspace{-0.25cm}
    \end{figure*}

\REV{In order to derive the approximate formula, we begin with a general expression for aperture efficiency as in \cite{Garcia_Matched_2020}:}
    \begin{equation} \label{eq:ApEff}
        \REV{\eta_{ap} = e_r\eta_a\eta_{s}\eta_t\eta_T,}
    \end{equation}

\noindent \REV{where $e_r$ is radiation efficiency and $\eta_a$ is achievement efficiency, both of which are approximately 1. $\eta_s$ is spillover efficiency, which is $\eta_s$=0.25 for all lens designs and frequencies considered in this work because the feed is an ideal 2D-isotropic line source with an $F/D=0.5$. $\eta_t$ is taper efficiency \cite{Stutzman-antenna}. $\eta_T$ is transmission efficiency determined by the lens impedance matching sections. We note that this is an effective lens transmission efficiency since the permittivity profile varies radially and thus each unit-cell has a unique transmission efficiency which combines to yield an overall lens transmission efficiency. With these assumptions we normalized the aperture efficiency, (\ref{eq:ApEff}), as:}
    \begin{equation} \label{eq:ApPrime}
        \REV{\eta_{ap}' =\frac{\eta_{ap}}{\eta_{s}}=\eta_T\eta_t.}
    \end{equation}

\REVNEW{Compared to $\eta_{ap}$, $\eta_{ap}'$ has divided out spillover efficiency ($\eta_s$), thus $\eta_{ap}'$ considers only the lens performance and not the feed. The use of $\eta_{ap}'$ also provides a more complete description of the frequency dependence of the matching efficiency compared to transmission efficiency $\eta_T$. While it may seem natural to characterize the matching performance of a lens antenna with $\eta_T$, in fact the matching has a small but noticeable affect on the amplitude distribution (e.g., a slightly higher reflection in the middle of the lens relative to the edge of the lens would result in a flattening of the amplitude distribution) and phase collimation of the lens. These effects are captured in the taper efficiency and included in $\eta_{ap}'$. Therefore we prefer to use $\eta_{ap}'$ to characterize the matching performance of a lens.}

\REV{There is another advantage to considering $\eta_{ap}'$ to characterize a GRIN lens---an approximate design expression $\widetilde{\eta_{ap}}$, can be derived. Suppose the phase collimation has been optimized through iteration such that taper efficiency is dominated by the amplitude distribution. Then the lens performance $\eta_{ap}'$ can be estimated by an effective $\widetilde{\eta_{ap}}$ which is a spatially averaged efficiency across the entire lens:}
    \begin{equation}    \label{ap_etaT}
        \REV{\widetilde{\eta_{ap}} \approx  \frac{1}{N}\sum_{n=1}^{N}(1-|\Gamma_n(f,\theta_{inc,n})|^2)^2\cdot\frac{1}{N}\sum_{n=1}^{N}\cos{\theta_{inc,n}}},
    \end{equation}
\noindent \REV{where $N$ is the number of rings in the GRIN lens (i.e., the number of unit-cells in the 2D cross section, here $N=250$). $\theta_{inc,n}$ is the incident angle from the point feed to the $n^{\textrm{th}}$ unit-cell. The first summation is the square of the effective transmission coefficient averaged over all $N$ unit cells, $\Bar{\eta_T}$. This quantity is squared to account for transmission through two tapers (bottom and top of lens). The second summation is an average taper efficiency, $\Bar{\eta_t}$, where phase is assumed to be perfectly collimated (and thus dominated by amplitude taper). $\Gamma_n(f)$ from (\ref{gammaKlop}) can be written in terms of design parameters and $\theta_{inc}$:}

  \begin{equation}
    \begin{split}
    \label{gammaKlop,inc}            \REV{\left|\Gamma(f,\theta_{inc})\right|=}&\REV{\Biggl|\Gamma_0\exp\left(-j\frac{2\pi f\sqrt{\epsilon_{eff(\theta_{inc})}}}{c_0}L\right)\cdot \Biggr.}\\
    &\REV{\Biggl.\frac{\cos\sqrt{\left(\frac{2\pi f\sqrt{\epsilon_{eff(\theta_{inc})}}}{c_0}L\right)^2-A^2}}{\cosh{A}}\Biggr|,}
    \end{split} 
    \end{equation} 

\noindent \REV{where $\epsilon_{eff}$ is a function of $\theta_{inc}$ and $\theta_{inc}=\arctan(r/F)$ is the incident angle of the unit-cell at radius $r$. Once $\theta_{inc}$ is determined, $\epsilon_{core}$ can be determined by the lens design algorithm in \cite{Garcia_Matched_2020} and $\epsilon_{eff}$ of that unit-cell can be derived by Algorithm\,\ref{alg:taper}\footnote{\REV{Formula (\ref{gammaKlop,inc}) uses the quasi-TEM assumption, which is valid within 45$^\circ$}}. 
Therefore, (\ref{ap_etaT}) can be rewritten as:}
   
    \begin{equation} \label{eq:ep_etaT}
    \REV{\widetilde{\eta_{ap}} \approx \Bar{\eta_T}\Bar{\eta_t}\approx \eta_T\eta_t = \eta_{ap}'.}
    \end{equation}

\REV{Referring back to Fig.\,\ref{fig:iteration loop}, $\widetilde{\eta_{ap}}$ (red traces) is largely invariant between iterative refining of the core permittivity profile but it is sensitive to the matching section specification. This indicates that $\widetilde{\eta_{ap}}$ is not dominated by the GRIN lens core profile so long as the lens design algorithm achieves approximate collimation in each iteration. Using $\widetilde{\eta_{ap}}$ as a target, iteration is performed to improve phase collimation and increase $\eta_{ap}'$ until its minima match $\widetilde{\eta_{ap}}$ in the final result. Therefore $\widetilde{\eta_{ap}}$ can be used as a design reference and provides a quick performance estimation.}



 \subsection{\REV{Performance Comparison}} \label{sec:Demo:Comparison}      
\REVNEW{Implementing the design and simulation methods above, a Klopfenstein taper matched lens (designed in Fig.\,\ref{fig:iteration loop}) and an unmatched lens are simulated to investigate the influence of taper operation. Figure\,\ref{fig:matVSunmat} shows the performance comparison between these two lenses. Based on the discussion of $\eta_{ap}'$ above, Fig.\,\ref{fig:matVSunmat}(a) shows the transmission efficiency $\eta_T$ and normalized aperture efficiency $\eta_{ap}'$ to be nearly the same which indicates the taper efficiency is almost unity (specifically in the passband $\eta_t$ is between $0.96$ and $0.98$). We also note that the shape of the curves is almost identical further confirming that $\eta_{ap}'$ is a good metric for comparing matching efficiency.}

\REVNEW{Referring again to Fig.\,\ref{fig:matVSunmat}(a), below the designed low-frequency cutoff of 11\,GHz, $\eta_{ap}'$ of both the Klopfenstein matched and the unmatched lenses oscillates rapidly due to multiple reflections. Above the taper cutoff frequency, however, $\eta_{ap}'$ of the matched lens improves significantly with much smaller oscillation due to the high transmission efficiency of the tapers. In order to further compare gain patterns of the two lenses five frequencies, 4.8 and 9.0\,GHz (below cutoff) and 16.0, 34.8, and 50.2\,GHz (within the passband) are indicated on the $\eta_{ap}'$ trace with colored markers and the associated $\eta_{ap}'$ values are included in the inset table.  The corresponding gain patterns are shown in Fig.\,\ref{fig:matVSunmat}(b) and the inset shows a zoom view of the pattern peaks. The solid traces are the Klopfenstein-matched lens and dotted traces are the unmatched lens. Gain patterns are each normalized to the peak gain of an aperture with $\eta_{ap}'=1$ at each frequency to make it simple to observe the reduction in aperture efficiency below cutoff. Below the cutoff frequency, we select frequencies where $\eta_{ap}'$ of both lenses are nearly the same and as a result the main lobes of two lenses are nearly identical. However, in the passband (above the cutoff frequency), gain maxima $G_o$ of the matched lens is around 1\,dB higher and the sidelobe levels (SLL) of the matched lens are better than those of the unmatched lenses. }

\REVNEW{Further comparison of $G_o$ and SLLs, defined as the ratio of the main beam to the first sidelobe, are shown in Fig.\,\ref{fig:matVSunmat}(c). Below the cutoff frequency the Klopfenstein-matched lens performs as an unmatched lens with nearly the same $G_o$. However, above the cutoff frequency the matched lens maintains a higher $G_o$ with around 1\,dB difference compared to that of the unmatched lens. As for SLL, below cutoff both lenses exhibit oscillatory SLL behavior versus frequency while in the passband the Klopfenstein-matched lens has improved SLL. The SLL of the unmatched lens oscillates to as low as 13.5\,dB across the entire band while the matched lens maintains a SLL above 15.5\,dB across the passband. We attribute the degradation in SLL for the unmatched lens to a flattening of the aperture amplitude distribution due to more pronounced reflections in the center of the lens. In general flat GRIN lenses of this type do not significantly alter the feed element amplitude taper as the field propagates through the lens (in part because as rays become more collimated the power density is maintained (rays do not diverge)). However, since the permittivity is highest in the center of the lens, the unmatched lens will have the largest reflection magnitude there and thus the higher fields in the center (due to the feed distribution) will be reduced due to reflections while the fields at the edges (where the permittivity is lowest) will have very little reflection. The result is a flattening of the amplitude distribution which results in a rise of the sidelobes (and corresponding drop in SLL as we have defined it here). Note that due to multiple reflections there are spot frequencies where $\eta_{ap}'$ is high and the SLL improves (large reflections in the middle of the lens are canceled due to advantageous multiple reflections) but in general, in order to maintain a consistent SLL versus frequency lens matching is necessary.}

\REVNEW{We note that the 2\,GHz periodic ripple in the Fig.\,\ref{fig:matVSunmat}(a) and (c) is caused by the resonance between lens lower and upper surfaces. Finally, note that $G_o$ is below 0\,dB at 4\,GHz and yet the inset pattern shows directionality. This is because the actual aperture efficiency is $\eta_{ap}\approx \eta_s \eta_{ap}'=0.25\times 0.65=-7.9\,$dB (accounting for the 25\% spillover efficiency of the 2D isotropic source used in the simulation). }
 
 \begin{figure*}[th]
\centering
\includegraphics[width=16cm]{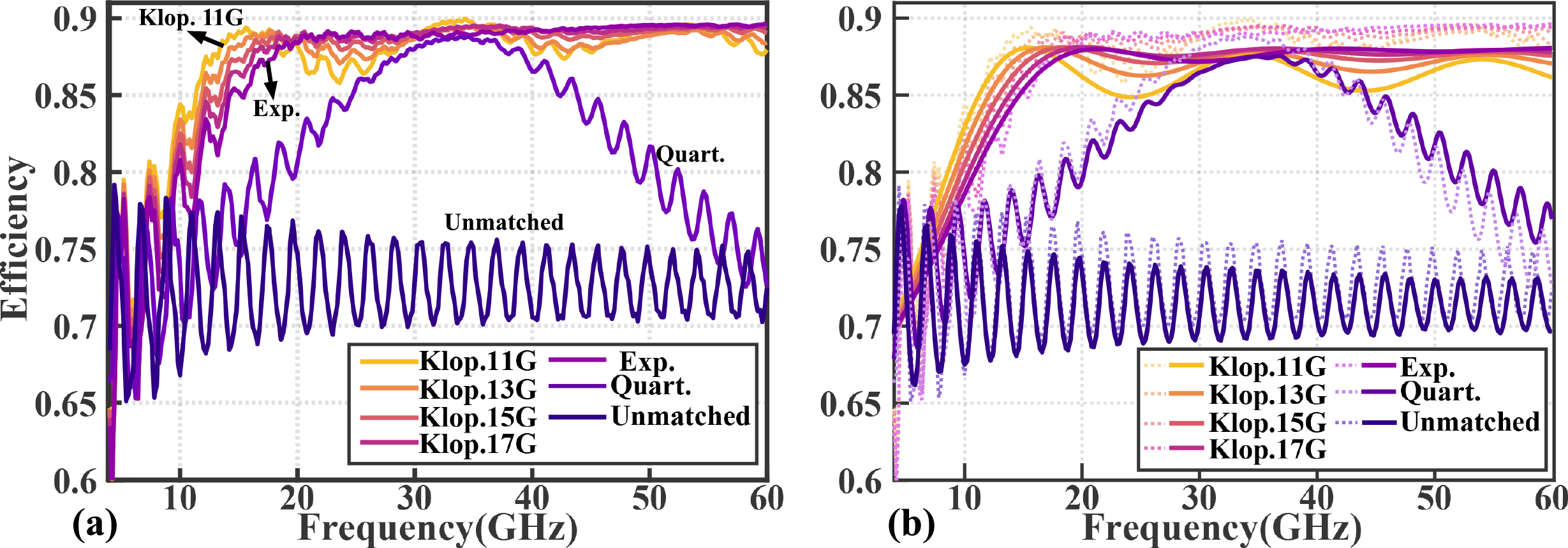}
\caption{\REV{(a) The comparison of $\eta_{ap}'$ for lenses with different types of tapers: Klopfenstein tapers (6 layers per side) with 4 different cutoff frequencies, Exponential tapers (6 layers), Quarter-wave matching sections (2 layers) and unmatched lens (0 layer). (b)The comparison of $\eta_{ap}'$ (dotted traces) and $\widetilde{\eta_{ap}}$ (solid traces) of all lenses plotted in (a).}}
\label{fig:multi_tapers}
\vspace{-0.25cm}
\end{figure*}
    
\REV{To further investigate the influence of taper types on overall lens performance Fig.\,\ref{fig:multi_tapers}(a) shows $\eta_{ap}'$ of GRIN lenses with Klopfenstein taper matching sections (labelled ``Klop.11G'' through ``Klop.17G''), exponential taper matching sections (as in  \cite{Garcia_Matched_2020}) (labelled ``Exp.''), quarter-wave matching sections (as in \cite{MaCui3_APL_2011,Qi_QuartLens_2013}) (labelled ``Quart.''), and unmatched lenses (as in \cite{Imbert_LTCClens_2017,mahmoud_unmatched_2014}) (labelled ``Unmatched''). Klopfenstein tapers with cutoff frequencies of 11, 13, 15, and 17\,GHz are included to show the tradeoff between cutoff frequency and passband efficiency. For the Klopfenstein tapers, based on formula (\ref{ga and band}), the larger the cutoff frequency, the smaller the passband ripples (and vice-versa). The traces show that the efficiency of an entire lens which uses many different Klopfenstein tapers (a unique one for each ring of the design) exhibits the expected tradeoff and has the same cutoff frequency as a single taper in isolation.} 

\REV{Because of the flexibility of Klopfenstein tapers, one can design lenses with similar passband efficiency as the Exponential matched lens but with lower cutoff frequency (see ``Klop.17G''). Here the 17\,GHz cutoff Klopfenstein taper achieves nearly identical passband efficiency but the low frequency cutoff is about 1--2\,GHz lower. A GRIN lens with quarter-wave matching sections was also designed for comparison. The matching sections were realized as a stack of two identical 0.030" layers so that the match frequency was within the passband of the taper-matched lenses (a single 0.030" quarter wave match has a match frequency at approximately 65\,GHz and is outside of our simulated band). At the exact match frequency the efficiency is similar to the Exponential and Klopfenstein matched lenses but degrades rapidly above and below this frequency. Finally, the unmatched lens is presented as a baseline with efficiency oscillating around approximately 72\%---the average efficiency is determined by the range of core permittivities used in the lens such that the unmatched efficiency would degrade if higher permittivities were used. The dramatic increase in efficiency of the taper-matched GRIN lenses compared to the unmatched case highlights the necessity of impedance matching for GRIN lenses. It is important to note that, in contrast to a one-dimensional taper which will exhibit a well-behaved and predictable frequency response, the GRIN lenses shown here have complicated responses which are the aggregate response of many unit-cells and have the added influence of a spatially displaced feed element (at the focal distance). Therefore we would not expect the GRIN lens efficiency to follow exactly the simpler trends of a single taper. Nevertheless, the overall trends hold which means that the taper design methods presented in this work can be used in the design of complete lenses.}

\REV{Figure\,\ref{fig:multi_tapers}(b) shows dotted semi-transparent $\eta_{ap}'$ versus solid $\widetilde{\eta_{ap}}$ from the final iteration for all lenses in Fig.\,\ref{fig:multi_tapers}(a). As frequency increases the non-idealities of the discretized tapers (as discussed in Section III) begin to distort the ideal taper response and this effect is exacerbated by the shorter taper lengths such that $\widetilde{\eta_{ap}}$ (which is based on idealized theory) deviates from $\eta_{ap}'$ (which is based on full-wave simulations). Nevertheless, $\eta_{ap}'$ and $\widetilde{\eta_{ap}}$ agree well within 5\% across the band. Notabely, $\widetilde{\eta_{ap}}$ shows the correct trends regarding cutoff frequency and passband efficiency indicating that the approximate $\widetilde{\eta_{ap}}$ are predictive and therefore useful for design. For the quarter-wave matched lens and the unmatched lens, $\Gamma(f,\theta_{inc})$ is derived by the wave transfer matrix method due to the lack of an analytical formula. As a result, the approximate $\widetilde{\eta_{ap}}$ includes the 2\,GHz ripple due to multiple reflections within the lens.}

\section{Conclusion}

In this work we have proposed a systematic design method for PUTs which minimally distort the ideal taper performance due to the use of non-commensurate lines. The design algorithm (Algorithm\,\ref{alg:taper}) and equivalent frequency response formula (\ref{ap_etaT}) are derived. The taper design method maintains key performance parameters of the chosen taper (e.g., $f_{c}$). We have explained the source of the nonidealities and proposed various means of mitigating them. The taper design method was validated \REV{by fabricating a nine layer Klopfenstein taper and characterizing it in a free-space (quasi-TEM) measurement setup across three waveguide bands. Finally, we presented an approximate expression for GRIN lens efficiency and designed GRIN lenses with several impedance matching sections, confirming the advantages of Klopfenstein taper matched GRIN lenses. The methods presented in this work enable designers to accurately predict the efficiency versus frequency of a single non-commensurate line taper as well as a complete GRIN lens comprising a large number of tapers to achieve optimal efficiency in high performance applications. The method allows for an accurate analysis of performance tradeoffs which was not previously possible.}


\bibliographystyle{ieeetr}
\bibliography{bib/TAP_taper}

\begin{IEEEbiography}[{\includegraphics[width=1in,height=1.25in,clip,keepaspectratio]{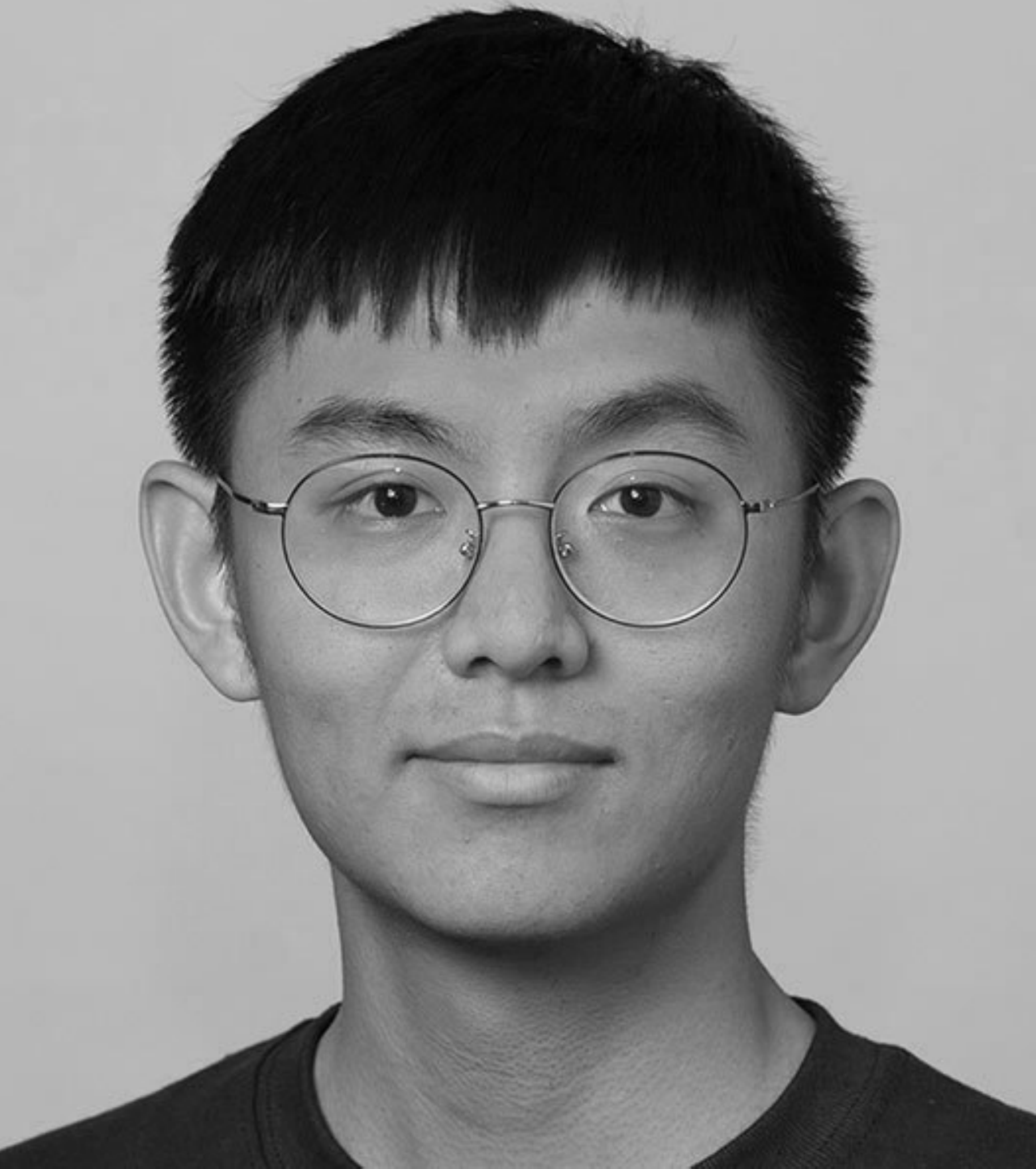}}]{Wei Wang} (Student Member, IEEE) is currently a Ph.D student in Electrical Engineering at the University of Notre Dame in Indiana, U.S.A. In 2019, he graduated with a B.S., Electrical Engineering from Nankai University in Tianjin, China. His research interests are gradient index(GRIN) lens antennas design and computational electromagnetics for GRIN lenses.
\end{IEEEbiography}

\begin{IEEEbiography}
[{\includegraphics[width=1in,height=1.25in,clip,keepaspectratio]{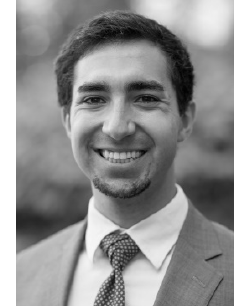}}]
{Nicolas~C.~Garcia} received the B.S. degree in electrical engineering from the University of Notre Dame in 2017. He is currently pursuing a Ph.D. at the same institution. His research interests focus on low-profile gradient index (GRIN) materials and their applications for emerging 5G and millimeter wave antennas and technologies.
\end{IEEEbiography}

\begin{IEEEbiography}[{\includegraphics[width=1in,height=1.25in,clip,keepaspectratio]{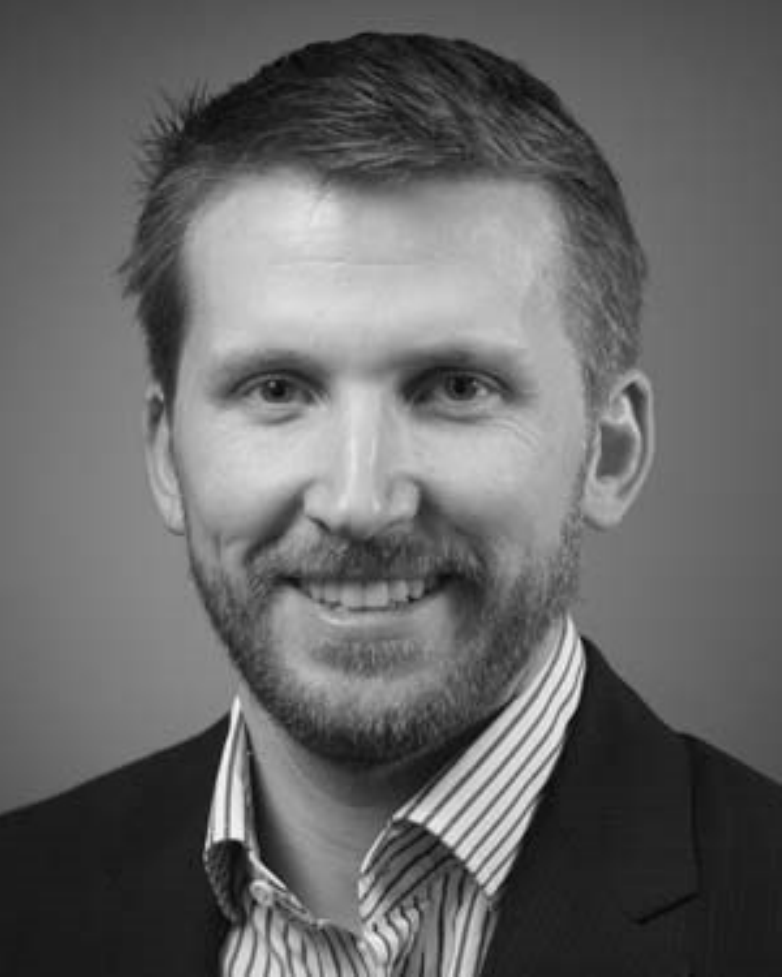}}]{Jonathan~D.~Chisum} (S'02--M'06--SM'17) received the Ph.D. in Electrical Engineering from the University of Colorado at Boulder in Boulder, Colorado USA, in 2011. 

From 2012 to 2015 he was a Member of Technical Staff at the Massachusetts Institute of Technology Lincoln Laboratory in the Wideband Communications and Spectrum Operations groups. His work at Lincoln Laboratory focused on millimeter-wave phased arrays, antennas, and transceiver design for electronic warfare applications. In 2015 he joined the faculty of the University of Notre Dame where he is currently an Assistant Professor of Electrical Engineering. His research interests include millimeter-wave communications and spectrum sensing with an emphasis on low-power and low-cost technologies. His group focuses on gradient index (GRIN) lenses for low-power millimeter-wave beam-steering antennas, nonlinear (1-bit) radio architectures for highly efficient communications and sensing up through millimeter-waves, as well as reconfigurable RF circuits for wideband distributed circuits and antennas. 

Dr. Chisum is a senior member of the IEEE, a member of the American Physical Society, and an elected Member of the U.S. National Committee (USNC) of the International Union or Radio Science's (URSI) Commission D (electronics and photonics). He is the past Secretary and current Vice-chair for USNC URSI Commission D: Electronics and Photonics.

\end{IEEEbiography}

\end{document}